\documentclass[11pt,a4paper]{article}
\pdfoutput=1
\usepackage{jheppub}
\usepackage{rotating}
\usepackage{array}
\usepackage{amsmath}
\usepackage{slashed}
\usepackage{booktabs}
\usepackage[pdftex,table]{xcolor}
\usepackage{mathtools}
\usepackage{empheq}
\usepackage{float}

\newcommand{\DM}{\chi}
\newcommand{\tD}{t_{\nu\text{-dec}}}
\newcommand{\TD}{T_{\nu\text{-dec}}}

\preprint{DESY 19-006}

\title{BBN constraints on the annihilation of MeV-scale dark matter}

\author{Paul Frederik Depta,}
\author{Marco Hufnagel,}
\author{Kai Schmidt-Hoberg}
\author{and Sebastian Wild}

\affiliation{DESY, Notkestra\ss e 85, D-22607 Hamburg, Germany}

\emailAdd{frederik.depta@desy.de}
\emailAdd{marco.hufnagel@desy.de}
\emailAdd{kai.schmidt-hoberg@desy.de}
\emailAdd{sebastian.wild@desy.de}

\abstract{
Thermal dark matter at the MeV scale faces stringent bounds from a variety of cosmological probes. Here we perform a detailed evaluation of BBN bounds on the annihilation cross section of dark matter with a mass 
1\,MeV\,$\lesssim m_\chi \lesssim 1\,$GeV.
For $p$-wave suppressed annihilations, constraints from BBN turn out to be significantly stronger than the ones from CMB observations, and are competitive with the strongest bounds from other indirect searches. We furthermore update the lower bound from BBN on the mass of thermal dark matter using improved determinations of primordial abundances. While being of similar strength as the corresponding bound from CMB, it is significantly more robust to changes in the particle physics model.
	}

\begin{document}
\maketitle

\section{Introduction}

Over the past few decades, theoretical work and experimental searches for dark matter (DM) have mainly focussed on particles with mass and
interactions set by the weak scale, motivated by the WIMP paradigm. The lack of an unambiguous sign of new physics at the LHC~\cite{CMS:2016pod,Aaboud:2016tnv,Kahlhoefer:2017dnp} as well as stringent 
bounds from direct DM searches~\cite{Aprile:2018dbl,Angloher:2015ewa,Agnese:2015nto}, however, motivate the study of alternative scenarios. In particular, DM below the GeV scale has attracted
a large amount of attention in the last few years as direct bounds are significantly weakened in this region of parameter space (see e.g.~\cite{Alexander:2016aln,Knapen:2017xzo,Darme:2017glc,Dutra:2018gmv,Matsumoto:2018acr}). For sub-GeV DM, additional light states
are typically required in order to obtain the correct DM relic abundance, leading to additional experimental signatures, which can be probed using astrophysical~\cite{Raffelt:1987yu,Krnjaic:2015mbs,Chang:2018rso} and cosmological~\cite{2016RvMP...88a5004C,Poulin:2016anj,Hufnagel:2017dgo,Hufnagel:2018bjp} observations as well as collider searches~\cite{Schmidt-Hoberg:2013hba,Dolan:2017osp,Kou:2018nap} and beam-dump experiments~\cite{Dolan:2014ska,Alekhin:2015byh,Dobrich:2015jyk}.
Furthermore, light mediating particles naturally lead to sizeable DM self-interactions, which may
be inferred through astrophysical observations and give a handle on these models independently of the couplings to SM states (see~\cite{Tulin:2017ara} for a review).

In this study we concentrate on cosmological implications of MeV-scale dark sectors with a particular focus on Big Bang Nucleosynthesis (BBN). 
Specifically, we consider the case of MeV-scale DM particles that can annihilate into the kinematically available Standard Model (SM) final states -- electrons, photons and neutrinos.
During standard BBN, such a scenario generically leads to an increase of the Hubble rate as well as to an injection of electromagnetic energy into the thermal bath of SM particles. Furthermore, if the DM freeze-out happens after neutrino decoupling, 
annihilations into electrons and photons (neutrinos) lead to a decrease (increase) of the neutrino-to-photon temperature ratio and thus to a modified effective number of neutrinos $N_\text{eff}$~\cite{Serpico:2004nm,Ho:2012ug,Boehm:2012gr,Steigman:2013yua,Boehm:2013jpa,Nollett:2013pwa,Nollett:2014lwa,Wilkinson:2016gsy}. In this paper, we present updated bounds on these scenarios by employing recent determinations of the deuterium and helium abundances~\cite{Olive:2016xmw}, and compare them to constraints from the Cosmic Microwave Background (CMB) using the latest data from Planck~\cite{Aghanim:2018eyx}.

We also point out that, for a sizeable branching into {\it both} channels, the DM-induced interactions of neutrinos with electrons can lower the neutrino decoupling temperature, potentially leading to an increase of $N_\text{eff}$\footnote{This has also been discussed in~\cite{Escudero:2018mvt} which appeared during the final stage of preparation for this work.}.

After the termination of nucleosynthesis, the products of residual DM annihilations may still have an effect on the abundances of light elements via photodisintegration processes~\cite{Reno:1987qw}. The resulting constraints on the annihilation cross section of DM have been derived in~\cite{Hisano:2009rc,Jedamzik:2009uy,Kawasaki:2015yya}, albeit only for masses above a few GeV. Extrapolating those bounds down to DM masses below $100\,$MeV is not possible due to the presence of photodisintegration thresholds. Furthermore, for these masses one has to account for potentially large deviations from the universal spectrum of photons originating from the electromagnetic cascade induced by the products of DM annihilation~\cite{Henning:2012rm,Poulin:2015woa,Poulin:2015opa,Hufnagel:2018bjp,Forestell:2018txr}. In this work we perform a dedicated study of BBN constraints on both the $s$- and $p$-wave annihilation cross section of MeV-scale DM, based on the computational techniques developed in~\cite{Hufnagel:2017dgo,Hufnagel:2018bjp}. In particular we show that the resulting bounds are especially competitive for the case of $p$-wave annihilations, as the average velocity of DM particles during the era of photodisintegration can be orders of magnitude larger than during recombination, leading to significantly weakened constraints from the CMB. We also compare our results to bounds derived from observations of gamma rays and charged cosmic rays.

This paper is organised as follows:~in section~\ref{sec:mod_hubblerate} we discuss the cosmological evolution of MeV-scale thermal DM and its impact on nuclear abundances via changes to the Hubble rate as well as to the photon and neutrino temperature. In section~\ref{sec:pdi} we then describe our method of calculating the non-thermal photon spectrum originating from the electromagnetic cascade induced by the residual annihilations of DM, and derive the resulting modifications to nuclear abundances from photodisintegration processes. In section~\ref{sec:results} we present the corresponding bounds from BBN and CMB, and compare our results to constraints from complementary observations. Finally, we conclude in section~\ref{sec:conclusions}. Additional material is provided in appendix~\ref{sec:additional_results} and \ref{app:pwavebounds}.

\section{Impact of MeV-scale DM on the abundances of light nuclei}
\label{sec:cosmo}

\subsection{Modifications to the Hubble rate and time-temperature relationships}
\label{sec:mod_hubblerate}

The presence of a DM particle $\DM$ with a mass $m_\DM$ at the MeV scale leads to modifications of the Hubble rate and the time evolution of the temperatures of the photon and neutrino bath, which in turn lead to a change in the predicted abundances of light nuclei. In the following we discuss the underlying formalism for two well-motivated particle physics scenarios: in the first case, (i), we assume that the interaction of DM with neutrinos is negligible, and thus only annihilations into electron-positron pairs and/or a pair of photons are relevant. This can be realized e.g.~in a scenario where the interaction is mediated by a Higgs-like scalar. In the second case, (ii), we consider equal branching ratios for the annihilation of DM into $e^+ e^-$ and $\nu_e \bar{\nu}_e$. This setup is naturally expected for models where DM couples to the first generation of SU(2)$_L$ lepton doublets. For $m_\DM < 2 m_e$, where the phase-space for annihilations into electron-positron pairs is closed, we assume equal rates for the annihilation into photon pairs and neutrinos, although such masses are in any case robustly excluded also for other choices of the branching ratios~\cite{Escudero:2018mvt}.

We denote the time of neutrino decoupling by $\tD$, and assume that neutrinos and photons are in full thermal contact for $t < \tD$, while they are completely decoupled at later times. Effects associated to non-instantaneous neutrino decoupling are sufficiently small for our purposes; see~\cite{Escudero:2018mvt} for a recent discussion. For scenario (i),  where no couplings between neutrinos and DM are present, $\tD$ is simply given by $t_{\nu \text{-}e^\pm\text{-dec}}$, i.e.~the time at which the SM scattering and annihilation processes involving neutrinos, electrons and positrons become inefficient. In the absence of additional particles contributing to the Hubble rate, the corresponding temperature is $T^{\text{(SM)}}(t_{\nu \text{-}e^\pm\text{-dec}}^{\text{(SM)}}) \simeq 2.3\,$MeV~\cite{Enqvist:1991gx}. In the presence of extra degrees of freedom, influencing the expansion rate and the time-temperature relation, we estimate  $t_{\nu \text{-}e^\pm\text{-dec}}$ as follows
\begin{equation}
T(t_{\nu \text{-}e^\pm\text{-dec}})^5/H(t_{\nu \text{-}e^\pm\text{-dec}}) \simeq T^{\text{(SM)}}(t_{\nu \text{-}e^\pm\text{-dec}}^{\text{(SM)}})^5/H^{\text{(SM)}}(t_{\nu \text{-}e^\pm\text{-dec}}^{\text{(SM)}}) \,,
\label{eq:nu_decoupling}
\end{equation}
observing that the relevant neutrino interaction rates scale as $\propto T^5$~\cite{Fradette:2017sdd,Hufnagel:2018bjp}.

For scenario (ii), where DM annihilates into both $e^+ e^-$ and $\nu_e \bar{\nu}_e$, DM can serve as a `bridge' for keeping neutrinos in equilibrium with the photon heat bath at times $t > t_{\nu \text{-}e^\pm\text{-dec}}$. Note that due to efficient oscillations, the flavor species of the final state neutrinos in the annihilation process is not important (see e.g.~the discussion in~\cite{Escudero:2018mvt}). The DM-induced equilibration of neutrinos with the photon heat bath becomes inefficient once $\Gamma_{\nu\text{-}\DM\text{-ann}}(t) \lesssim H(t)$, where $\Gamma_{\nu\text{-}\DM\text{-ann}}(t)$ is the rate of DM annihilations into $\nu_e \bar{\nu}_e$ (which by assumption in scenario (ii) is the same as the rate of DM annihilations into $e^+ e^-$). Defining $t_{\nu\text{-}\DM\text{-dec}}$ via $\Gamma_{\nu\text{-}\DM\text{-ann}}(t_{\nu\text{-}\DM\text{-dec}}) \simeq H(t_{\nu\text{-}\DM\text{-dec}})$, we can then estimate the time of neutrino decoupling via
\begin{align}
\tD \simeq \max\left[ t_{\nu \text{-}e^\pm\text{-dec}}, t_{\nu\text{-}\DM\text{-dec}} \right] \,.
\label{eq:def_tD}
\end{align}
Clearly, this is only an approximation to the more complex mechanism underlying the decoupling of neutrinos, but it is nevertheless sufficient for qualitatively understanding the impact on BBN and CMB observables. An alternative approach based on the consideration of energy transfer rates has been recently advocated in~\cite{Escudero:2018mvt} with results presented in the limiting cases $\left<\sigma v\right>_{\chi\chi \rightarrow e^+ e^-} \gg \left<\sigma v\right>_{\chi\chi \rightarrow \nu_e \bar{\nu}_e}$ and $\left<\sigma v\right>_{\chi\chi \rightarrow e^+ e^-} \ll \left<\sigma v\right>_{\chi\chi \rightarrow \nu_e \bar{\nu}_e}$.

Having determined the time of neutrino decoupling $\tD$, let us now turn to the calculation of the common temperature $T(t) = T_\nu(t)$  of the photon and neutrino bath for $t < \tD$. For both scenarios (i) and (ii), the entropy density of all particles in thermal equilibrium is given by\footnote{Here we assume that DM is in equilibrium with the heat bath at temperatures where $F_s (m_\DM/T) \sim \mathcal{O}(1)$. This requirement is satisfied for the DM annihilation cross section that corresponds to the observed relic density.}
\begin{align}
s(T) \big|_{t < \tD}^{\text{(i), (ii)}} = \frac{2 \pi^2}{45} \left( g_s^{(T_\nu = T)} (T) + g_\DM F_s (m_\DM/T)\right) T^3 \,.
\label{eq:s_BeforeNuDec}
\end{align}
Here, $g_s^{(T_\nu = T)} (T)$ denotes the effective number of SM degrees of freedom for $T_\nu = T$, $g_\DM$ is the number of internal degrees of freedom of the DM particle, and $F_s (x)$ is given by
\begin{align}
F_s(x) = \frac{15}{4 \pi^4} \int_{x}^\infty \text{d}u \, \frac{3 u^2 (u^2 - x^2)^{1/2} + (u^2 - x^2)^{3/2}}{e^u \pm 1} \,,
\end{align}
with the $+$ ($-$) sign corresponding to fermionic (bosonic) DM. Entropy conservation implies $\dot{s} + 3 H s = 0$, and after inserting Eq.~(\ref{eq:s_BeforeNuDec}) we obtain $T(t)$ from the solution of the resulting differential equation.
After neutrino decoupling, entropy is conserved separately in both the photon and the neutrino sector. 

In scenario (i), in which DM only annihilates into $e^+ e^-$ and/or $\gamma \gamma$, DM can still be in equilibrium with the photon heat bath even after neutrino decoupling. Thus, the entropy density of all particles that are still in equilibrium with the photons is given by
\begin{align}
s_\gamma(T)\big|_{t > \tD}^{\text{(i)}} = \frac{2 \pi^2}{45} \left( g_s^{\text{(vis)}} (T) + g_\DM F_s (m_\DM/T)\right) T^3 \,,
\end{align}
with $g_s^{\text{(vis)}}(T)$ being the effective number of degrees of freedom of the photons, electrons and positrons. The time evolution of $T(t)$ again follows from solving the equation of entropy conservation in the photon sector, $\dot{s}_\gamma + 3 H s_\gamma = 0$, while the entropy conservation in the neutrino sector can be used to obtain the corresponding neutrino temperature
\begin{align}
\frac{T_\nu(T)}{T} \bigg|_{t > \tD}^{\text{(i)}} = \left(\frac{g_s^\text{(vis)} (T) + g_\DM F_s (m_\DM/T)}{g_s^\text{(vis)} (\TD) + g_\DM F_s (m_\DM/\TD)}\right)^{1/3} \,,
\end{align}
with $\TD \equiv T(\tD)$.

In scenario (ii), in which DM annihilates into $e^+ e^-$ and $\nu_e \bar{\nu}_e$, the definition of $\tD$ in Eq.~(\ref{eq:def_tD}) implies that the DM particles are neither in equilibrium with the photon nor with the neutrino bath for $t > \tD$, and hence do not contribute to the entropy degrees of freedom. We thus simply have
\begin{align}
s_\gamma(T)\big|_{t > \tD}^{\text{(ii)}} = \frac{2 \pi^2}{45} g_s^{\text{(vis)}} (T) T^3 \,,
\end{align}
and determine $T(t)$ again from $\dot{s}_\gamma + 3 H s_\gamma = 0$. The neutrino temperature in this scenario is then given by
\begin{align}
\frac{T_\nu(T)}{T} \bigg|_{t > \tD}^{\text{(ii)}} = \left(\frac{g_s^\text{(vis)} (T)}{g_s^\text{(vis)} (\TD)}\right)^{1/3} \,.
\end{align}

Finally, we compute the expected abundances of ${}^1$H, D, ${}^3$He and ${}^4$He by using a modified version of \textsc{AlterBBN~v1.4}~\cite{Arbey:2011nf,Arbey:2018zfh}. Similar to the procedure described in~\cite{Hufnagel:2018bjp}, we replace the built-in functions for the temperatures $T(t)$ and $T_\nu(t)$ as well as the Hubble rate $H(t)$ by the results of our calculations outlined above. The baryon-to-photon ratio is fixed to the value $\eta = 6.1\times 10^{-10}$~\cite{Ade:2015xua} at the time of recombination, and consistently propagated backwards in time using entropy conservation.

\subsection{Photodisintegration induced by residual annihilations of DM}
\label{sec:pdi}

After freeze-out, the comoving number density of DM stays approximately constant. Nevertheless, residual annihilations with a rate $\propto n_\DM^2 \times \langle \sigma v \rangle$ are still taking place, and can lead to modifications in the predicted abundances of light nuclei. Specifically, if DM annihilates into $e^+ e^-$ and/or $\gamma \gamma$, these annihilation products induce an electromagnetic cascade, leading to a non-thermal spectrum of photons which can potentially destroy light elements, e.g.~via $d(\gamma,p)n$. It is well known that these photodisintegration  processes are only effective at $t \gtrsim 10^4\,$s, as for smaller injection times the resulting cascade spectrum falls below the disintegration threshold of all light nuclei~\cite{Cyburt:2002uv}.

For sufficiently large DM masses, the photons originating from the cascade process follow a universal spectrum, with a normalisation depending only on the total amount of injected energy~\cite{Cyburt:2002uv, Kawasaki:1994sc}. However, as pointed out in~\cite{Serpico:2015blj}, for the ranges of injection energies relevant to this study, the universal spectrum is typically not applicable. We thus compute the energy spectrum $f_X(E)$ of the particles $X \in \{ \gamma, e^-, e^+ \}$ from scratch by solving the relevant cascade equations~\cite{Kawasaki:1994sc,Jedamzik:2006xz,Hufnagel:2018bjp}
\begin{align}
f_X(E) = \frac{1}{\Gamma_X(E)} \left( S_X(E) + \int_{E}^{\infty} \text{d} E' \sum_{X'} \left[ K_{X' \to X} (E, E') f_{X'}(E') \right] \right)\,,
\label{eq:fX_recursive}
\end{align}
with the total interaction rate $\Gamma_X(E)$, the corresponding differential interaction rate for scattering and/or conversion $K_{X' \to X} (E, E')$ as well as the source term $S_X(E)$. Note that the dependence on the temperature $T$ has been suppressed to avoid clutter. For the case of residual DM annihilations, the source term is given by
\begin{align}
S_X(E) = S_X^{(0)} \delta(E-m_\DM) + S_X^\text{(FSR)}(E) \,.
\label{eq:SXE_definition}
\end{align}
Here, the first term corresponds to the monochromatic energy injection of the particle $X$ induced by the annihilation of non-relativistic self-conjugate DM particles with mass $m_\DM$:
\begin{align}
S_{e^-}^{(0)} = S_{e^+}^{(0)} &= \frac12 n_\DM^2 \left< \sigma v \right>_{\DM \DM \rightarrow e^+ e^-}\,, \label{eq:Se0}\\
S_{\gamma}^{(0)} &= n_\DM^2 \left< \sigma v \right>_{\DM \DM \rightarrow \gamma \gamma} \,. \label{eq:Sgamma0}
\end{align}
Note that the annihilation into neutrinos does not lead to an electromagnetic cascade and thus does not further appear in the discussion of photodisintegration. Following~\cite{Forestell:2018txr,Mardon:2009rc, Birkedal:2005ep}, the second term in Eq.~\eqref{eq:SXE_definition} accounts for the final state radiation (FSR) of photons for DM annihilating into $e^+ e^-$:
\begin{align}
S_\gamma^\text{(FSR)}(E) \, = \,\,\, &\frac{n_\DM^2 \left< \sigma v \right>_{\DM \DM \rightarrow e^+ e^-}}{2m_\DM} \,\times \nonumber\\
&\quad\frac{\alpha}{\pi} \frac{1+(1-x)^2}{x}\ln\left( \frac{4m_\DM^2(1-x)}{m_e^2} \right) \times \Theta\left( 1 - \frac{m_e^2}{4m_\DM^2} - x \right) \,,
\label{eq:SFSR}
\end{align}
with $x=E/m_\DM$.

Since the DM particles are non-relativistic for all temperatures relevant for photodisintegration, we can expand the thermally averaged annihilation cross section appearing in Eqs.~(\ref{eq:Se0})$-$(\ref{eq:SFSR}) in powers of the relative velocity $v_\text{rel}$:
\begin{align}
\left< \sigma v\right> \simeq a + b \left< v_\mathrm{rel}^2 \right>  \,.
\end{align}
In the following, we will consider both the case of unsuppressed $s$-wave annihilations with $\left< \sigma v \right>$ being dominated by $a$ as well as a scenario with $p$-wave annihilating DM, in which case the normalisation of the source terms depends only on $b$. In the latter case, the thermal average of the squared relative velocity of two annihilating DM particles is given by
\begin{align}
\left< v_\mathrm{rel}^2 \right> \simeq \frac{6T_{\DM}(T)}{m_\DM}\,.
\label{eq:vrel2}
\end{align}

\begin{figure}
	\centering
	\includegraphics[width=0.8\textwidth]{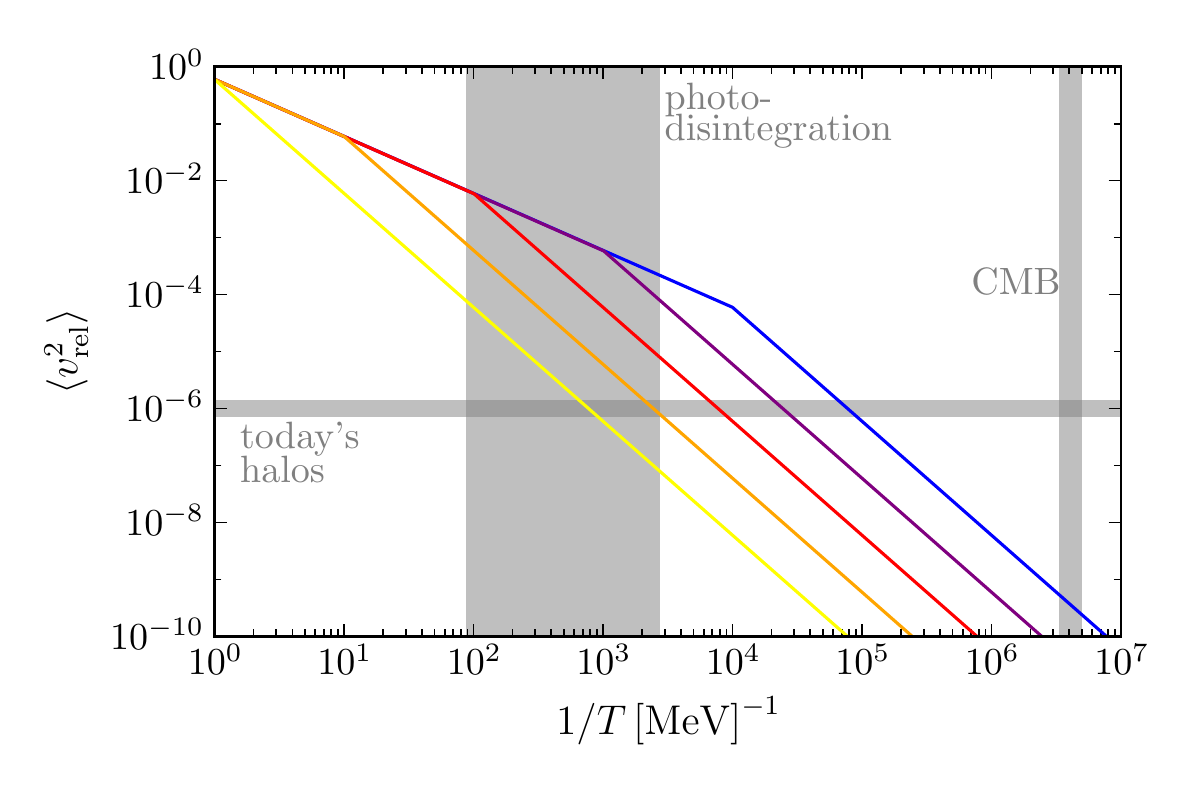}
	\caption{Thermally averaged squared velocity of DM for $m_\DM = 10\,$MeV and different choices of the kinetic decoupling temperature $T^\text{kd} =\,$100~eV, 1~keV, 10~keV, 100~keV and 1~MeV (from blue to yellow). The gray shaded regions roughly indicate the range of temperatures or velocities relevant to photodisintegration, CMB observables and DM annihilation in typical present-day halos.}
	\label{fig:v2DM}
\end{figure}

The temperature of DM particles after chemical decoupling $T_\DM(T)$ entering this expression critically depends on $T^\text{kd}$, the temperature at which DM kinetically decouples from the photon heat bath~\cite{Kolb:1990vq}:
\begin{align}
T_\DM(T) = 
\begin{cases}
    T& \text{if } T> T^{\text{kd}}\,,\\
    T^\text{kd}  R( T^\text{kd})^2/R(T)^2              & \text{if } T < T^\text{kd}\,,
\end{cases}
\label{eq:TDM_T_R}
\end{align}
where $R(T)$ is the scale factor at a given temperature. In the following we pursue a model-independent approach by treating the kinetic decoupling temperature as a free parameter, noting that values down to $T^\text{kd} \gtrsim 100\,$eV are consistent with Lyman-$\alpha$ forest measurements~\cite{Bringmann:2016ilk}. We show the resulting temperature dependence of $\left< v_\mathrm{rel}^2 \right>$ in Fig.~\ref{fig:v2DM} for different values of the kinetic decoupling temperature, choosing $m_\DM = 10\,$MeV for concreteness. Clearly, for a wide range of kinetic decoupling temperatures, the suppression of the annihilation cross section by the square of the DM relative velocity is much less severe during photodisintegration than at recombination. The resulting values of $\left< v_\mathrm{rel}^2 \right>$ are also typically larger than the ones entering indirect detection probes of DM annihilating in e.g.~the Milky Way or dwarf galaxies. As we will see in section~\ref{sec:results}, this implies BBN bounds on the $p$-wave annihilation cross section which are much stronger than the corresponding limits from the CMB, and also very competitive with other indirect detection probes.

After calculating the source terms $S_X(E)$ by using the formalism described above, we solve the coupled cascade equations~(\ref{eq:fX_recursive}) following the method presented in~\cite{Hufnagel:2018bjp}. To this end, we take into account all relevant processes contributing to the scattering and conversion of photons, electrons and positrons, including double photon pair creation, photon-photon scattering, Bethe-Heitler pair creation, Compton scattering, and inverse Compton scattering\footnote{See~\cite{Hufnagel:2018bjp} for a collection of the corresponding expressions for the rates $\Gamma_X$ and $K_{X'\rightarrow X}$.}. In order to obtain the modifications to the abundances of light nuclei induced by the photodisintegration processes, it is important to note that at the time where photodisintegration becomes relevant ($t \gtrsim 10^4\,$s) nucleosynthesis has already terminated. For a given point in parameter space we can therefore take the abundances calculated as described in section~\ref{sec:mod_hubblerate}, and then evolve those according to~\cite{Cyburt:2002uv, Poulin:2015opa}
\begin{align}
\left( \frac{\text{d} T}{\text{d} t} \right) \frac{\text{d} Y_X}{\text{d} T} & = \sum_{N_i} Y_{N_i} \int_{0}^{\infty}\text{d} E\; f_\gamma(E)\sigma_{\gamma + N_i \rightarrow X}(E) \nonumber \\
&\;\, - Y_X \sum_{N_f} \int_{0}^{\infty}\text{d} E\; f_\gamma(E)\sigma_{\gamma + X \rightarrow N_f}(E) \,,
\label{eq:dYdT}
\end{align}
where $Y_X = n_X/n_b$ and $X \in \{ p, n, {}^2\text{H}, {}^3\text{H}, {}^3\text{He}, {}^4\text{He}, \dots \}$. Details on the relevant photodisintegration cross sections as well as on the method of solving Eq.~(\ref{eq:dYdT}) can be found in~\cite{Hufnagel:2018bjp}. 

\section{Constraints from BBN and CMB}
\label{sec:results}

\subsection{Comparison with observations}

We confront the predicted abundances of light nuclei calculated as described in the previous section to the following set of observed abundance ratios\footnote{Note that throughout this work we employ the $2\sigma$ upper bound on the abundance ratio ${}^3\text{He/D}$, which is generally accepted to be a robust cosmological probe \cite{Geiss2003,Kawasaki:2004qu,Ellis:2005ii}. Using the primordial value of ${}^3\text{He}/{}^1\text{H}$  inferred in~\cite{Bania:2002yj} would result in a slightly stronger bound. However, this value has been argued to be subject to sizeable astrophysical uncertainties and whether this should be used to constrain early universe cosmology is debated in the literature~\cite{VangioniFlam:2002sa,Kawasaki:2004qu,Olive:2016xmw}.
We also do not consider constraints from lithium, for which the observed abundance is in disagreement with the expectation in the SM~\cite{Fields:2011zzb}, but also subject to large astrophysical uncertainties~\cite{Korn:2006tv}.}:
\begin{align}
& \mathcal{Y}_\text{p} \quad & (2.45 \pm 0.04) \times 10^{-1} \label{eq:Yp_abundance} \; \,,\quad\text{\cite{Olive:2016xmw}}\\
& \text{D}/{}^1\text{H} \quad & (2.53 \pm 0.04) \times 10^{-5} \;\,,\quad\text{\cite{Olive:2016xmw}} \label{eq:D_abundance}\\
& {}^3\text{He}/\text{D} \quad & (8.3 \pm 1.5) \times 10^{-1} \label{eq:3HeH_abundance} \;\,.\quad\text{\cite{Geiss2003}}
\end{align}

Theoretical uncertainties associated to the nuclear rates entering our calculation are taken into account by running \textsc{AlterBBN~v1.4} in three different modes corresponding to high, low and central values for the relevant rates. We then derive $95\%\,$C.L. bounds by combining the observational and theoretical errors as described in more detail in~\cite{Hufnagel:2018bjp}.

Thermal MeV-scale DM is also constrained by CMB observations. As explained in section~\ref{sec:mod_hubblerate}, depending on the annihilation channels the freeze-out of DM can lead to a delayed thermal decoupling of neutrinos, as well as to an enhancement or a suppression of the neutrino temperature relative to the photon temperature. This implies a non-standard value of the effective number of neutrinos at recombination ($t = t^\text{rec}$):
\begin{align}
N_\text{eff}^\text{(CMB)} = N_\nu \left( \frac{T_\nu(t^\text{rec})}{T(t^\text{rec})} \right)^4 \left( \frac{11}{4} \right)^{4/3} \;\,.
\label{eq:NeffCMB_def}
\end{align}
In the following, we set the number of SM neutrinos to $N_\nu = 3.032$ in order to match the precise SM result $N_\text{eff}^\text{(CMB)} = 3.046$ obtained by taking into account effects associated to non-instantaneous neutrino decoupling and QED corrections to the electron mass~\cite{Mangano:2005cc,deSalas:2016ztq}. For constraining $N_\text{eff}^\text{(CMB)}$ we employ the latest results from Planck~\cite{Aghanim:2018eyx}, using the combination of TT, TE, EE+lowE, lensing, and BAO data. In view of the partial degeneracy in the experimental determination of $N_\text{eff}^\text{(CMB)}$ and $\mathcal{Y}_\text{p}$ at the time of recombination, we consistently calculate both observables for each scenario, and confront them to the 95\% C.L.~region in the plane spanned by $N_\text{eff}^\text{(CMB)}$ and $\mathcal{Y}_\text{p}$ (shown in Fig.~41 of~\cite{Aghanim:2018eyx}).

\subsection{Bounds on the mass of thermal DM from BBN and CMB}
\label{sec:res_mass_bound}

\begin{figure}
	\centering
	\hspace*{-2.5cm}
	\includegraphics[width=0.60\textwidth]{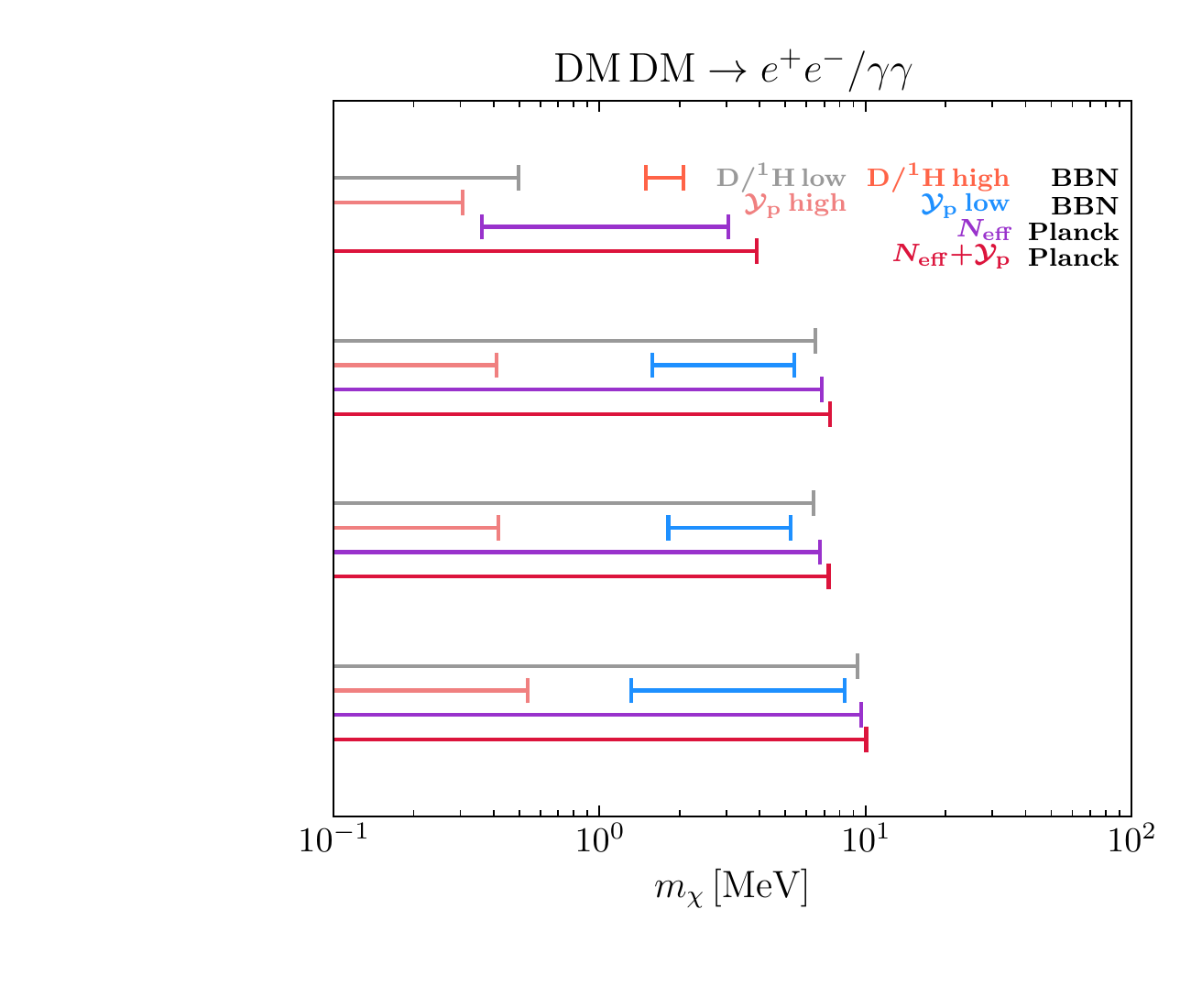}\hspace*{-0.42cm}
	\includegraphics[width=0.60\textwidth]{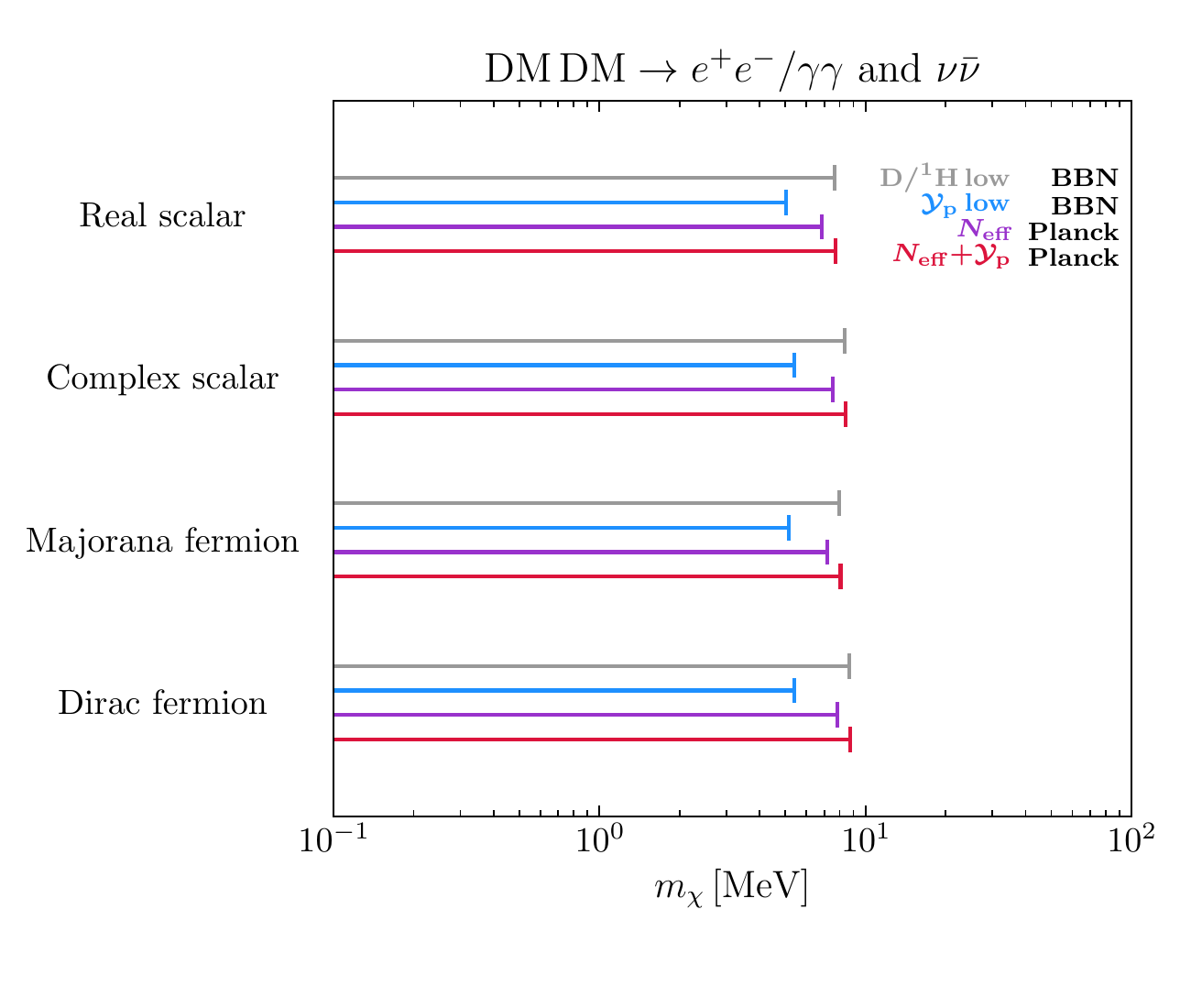}
	\vspace{-0.75cm}
	\caption{Bounds on the mass of a thermal DM particle annihilating into $e^+ e^-$ and/or $\gamma \gamma$ (left panel), or into electromagnetic final states and neutrinos with the same branching ratio (right panel). The region excluded by (under-) overproduction of D/$^1$H is shown in gray (orange), and similarly for $\mathcal{Y}_\text{p}$ in pink and blue, respectively. Constraints from the CMB are shown in purple and dark red (see text for details).} 
	\label{fig:bounds_thermalDMMmass}
\end{figure}

In Fig.~\ref{fig:bounds_thermalDMMmass} we show the range of DM masses excluded by different combinations of observables, assuming a thermal annihilation cross section of $\langle \sigma v \rangle \simeq 4 \times 10^{-26}\,\text{cm}^3/\text{s}$ ($8 \times 10^{-26}\,\text{cm}^3/\text{s}$) for self-conjugated (not self-conjugated) DM~\cite{Steigman:2012nb}. The left panel corresponds to a DM particle annihilating exclusively into $e^+ e^-$ and/or $\gamma \gamma$, while the right panel assumes equal branching ratios into electromagnetic final states and neutrinos. For each type of DM particle, the constraints from BBN are shown in gray and orange for D/$^1$H and in pink and blue for $\mathcal{Y}_\text{p}$; ${}^3$He does not appear as it leads to less stringent constraints on this scenario. The CMB bounds corresponding to modified values of $N_\text{eff}^\text{(CMB)}$ and $\mathcal{Y}_\text{p}$ are shown in purple and dark red, respectively. In the former case, we only employ data from Planck, lensing and BAO, implying a rather large range of allowed values for the helium abundance~\cite{Aghanim:2018eyx}. In the latter case, we additionally impose the constraint from the direct observation of $\mathcal{Y}_\text{p}$, corresponding to Eq.~(\ref{eq:Yp_abundance}).

With the exception of a real scalar annihilating into $e^+ e^-$ or $\gamma \gamma$, we find that the BBN bounds from D/$^1$H are nearly identical to the corresponding constraints from the CMB: depending on the DM type and annihilation channel, both observations rule out thermal DM with a mass $m_\DM \lesssim (7 -10)\,$MeV. Crucially, this bound applies to both $s$- and $p$-wave annihilating DM. It is important to stress that the constraint obtained from BBN is significantly more robust. Specifically, for the scenario of DM annihilating exclusively into electromagnetic final states, the CMB constraint arises from a value of $N_\text{eff}^\text{(CMB)}$ \emph{smaller} than the observed one, which can quite easily be compensated by invoking additional (dark) degrees of freedom contributing to the energy density. On the other hand, the BBN bound mainly arises from the additional contribution of the DM particle to the Hubble rate, which would only get stronger in the presence of dark radiation. The BBN bound can therefore be considered model-independent unlike the constraint from the CMB.

For DM annihilations into electromagnetic final states, the CMB bounds shown in the left panel of Fig.~\ref{fig:bounds_thermalDMMmass} are in good agreement with~\cite{Escudero:2018mvt}, while they are slightly stronger than the ones derived in~\cite{Boehm:2013jpa,Nollett:2013pwa} due to the use of the most recent Planck data in this work. On the other hand, our BBN bounds for DM annihilating into $e^+ e^-/\gamma \gamma$ are significantly stronger than in~\cite{Boehm:2013jpa,Nollett:2013pwa}, mostly due to the more precise measurement of D/$^1$H (see Eq.~(\ref{eq:D_abundance})). Finally, the CMB and BBN bounds on the scenario of DM annihilating into $e^+ e^-/\gamma \gamma$ and $\nu_e \bar{\nu}_e$ with equal branching ratios are derived for the first time in this work.

\subsection{Bounds on the annihilation cross section of DM from photodisintegration}
\label{sec:results_pdi}

\begin{figure}
	\centering
	\includegraphics[width=0.495\textwidth]{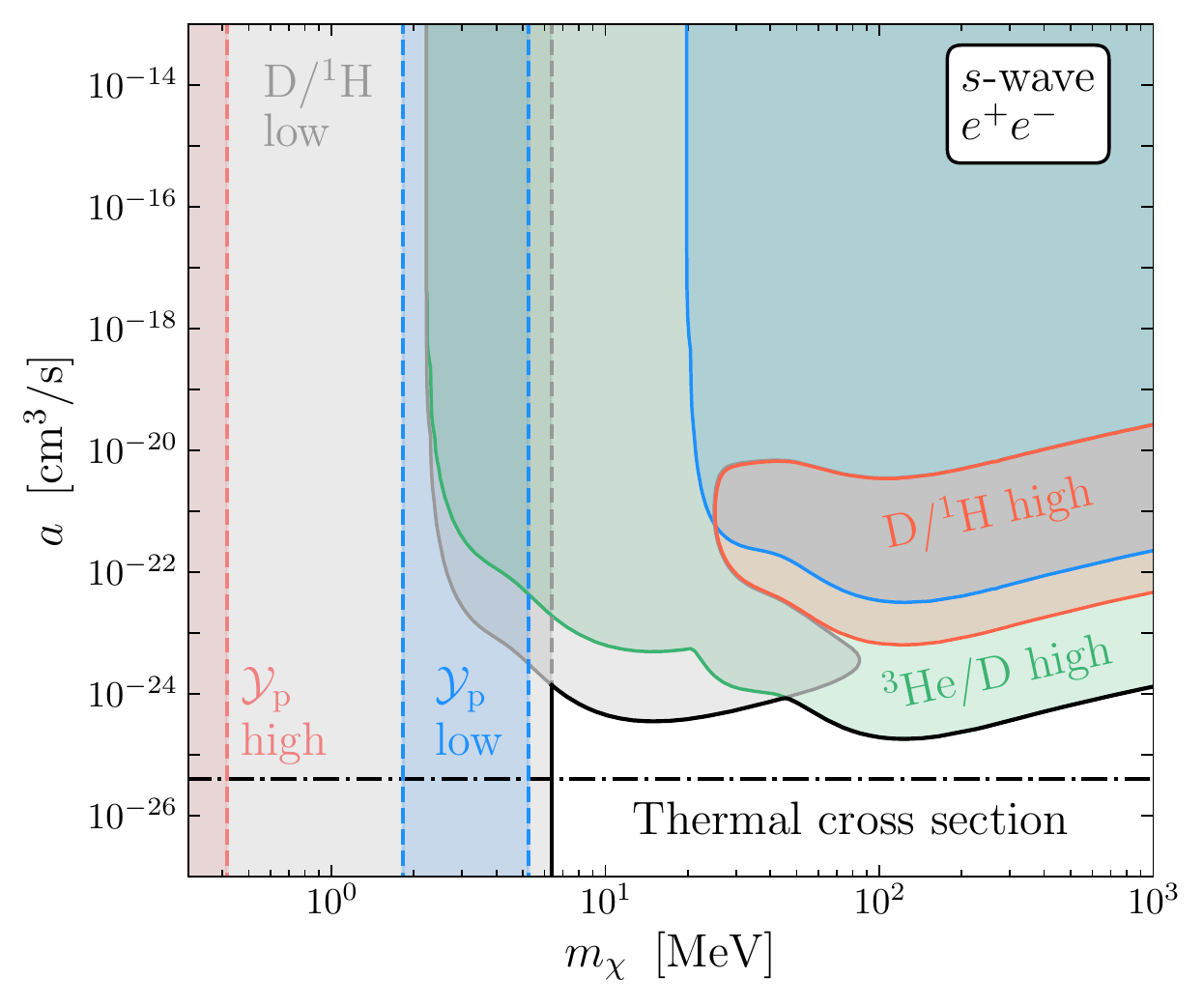}
	\includegraphics[width=0.495\textwidth]{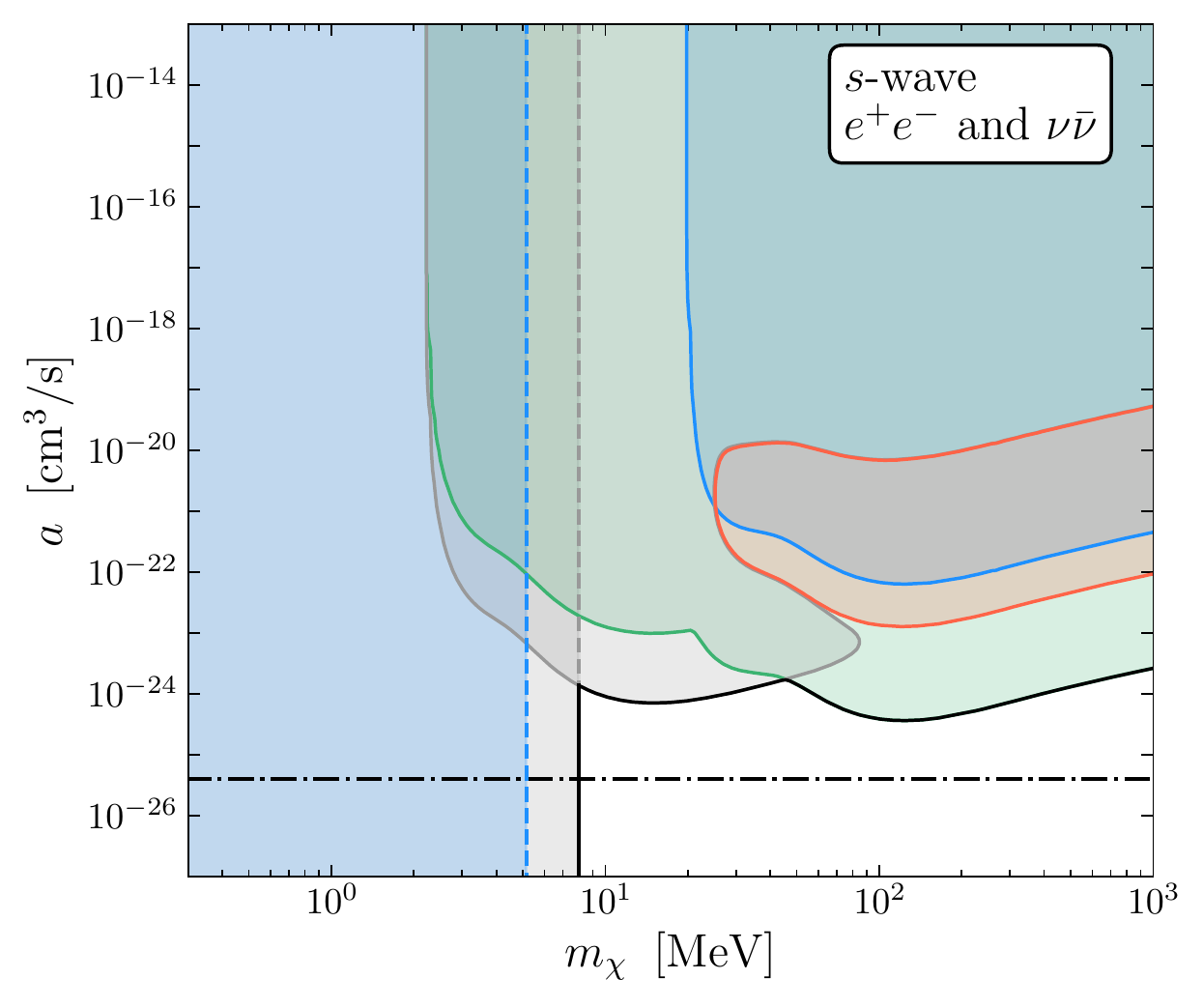}
	\includegraphics[width=0.495\textwidth]{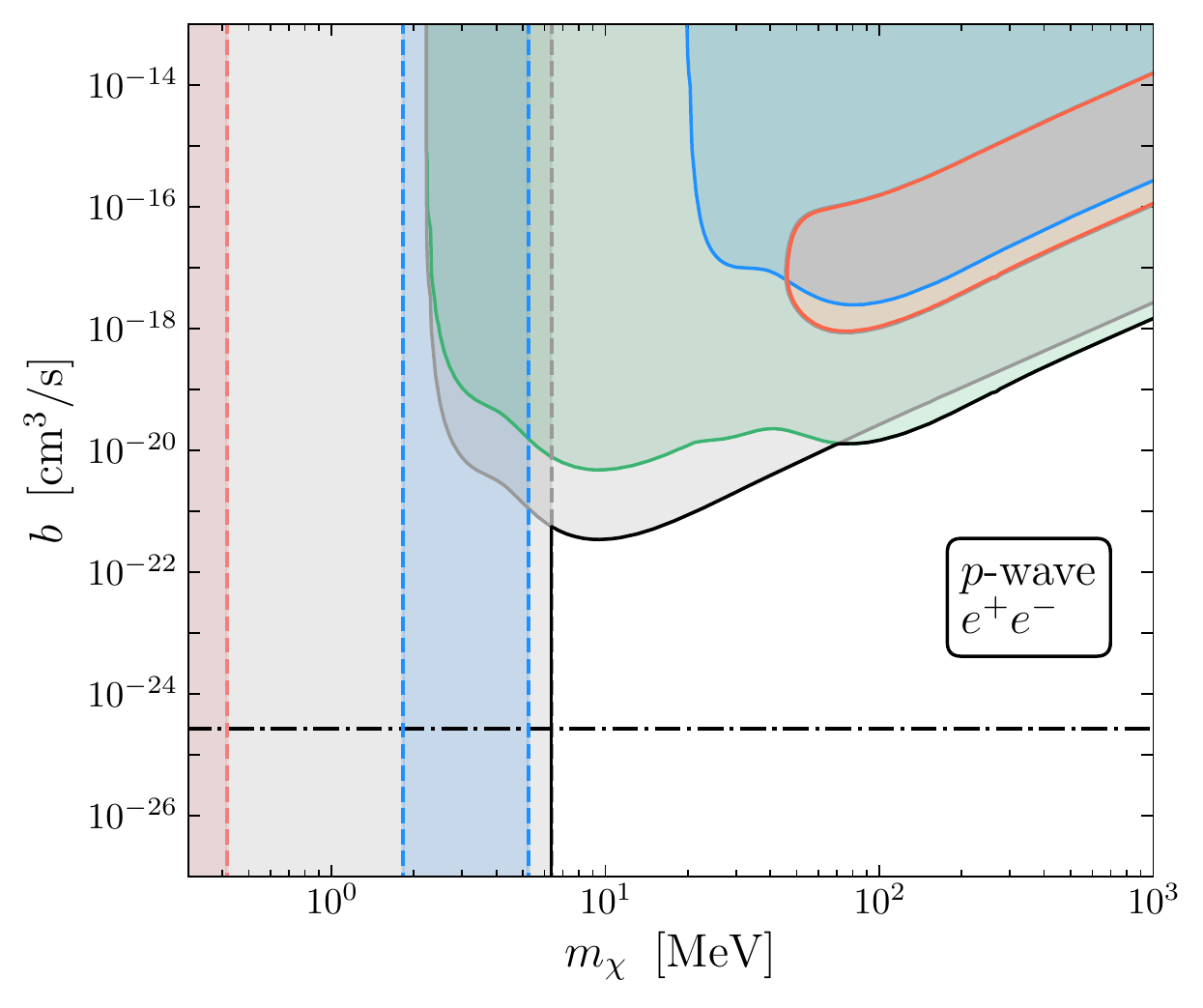}
	\includegraphics[width=0.495\textwidth]{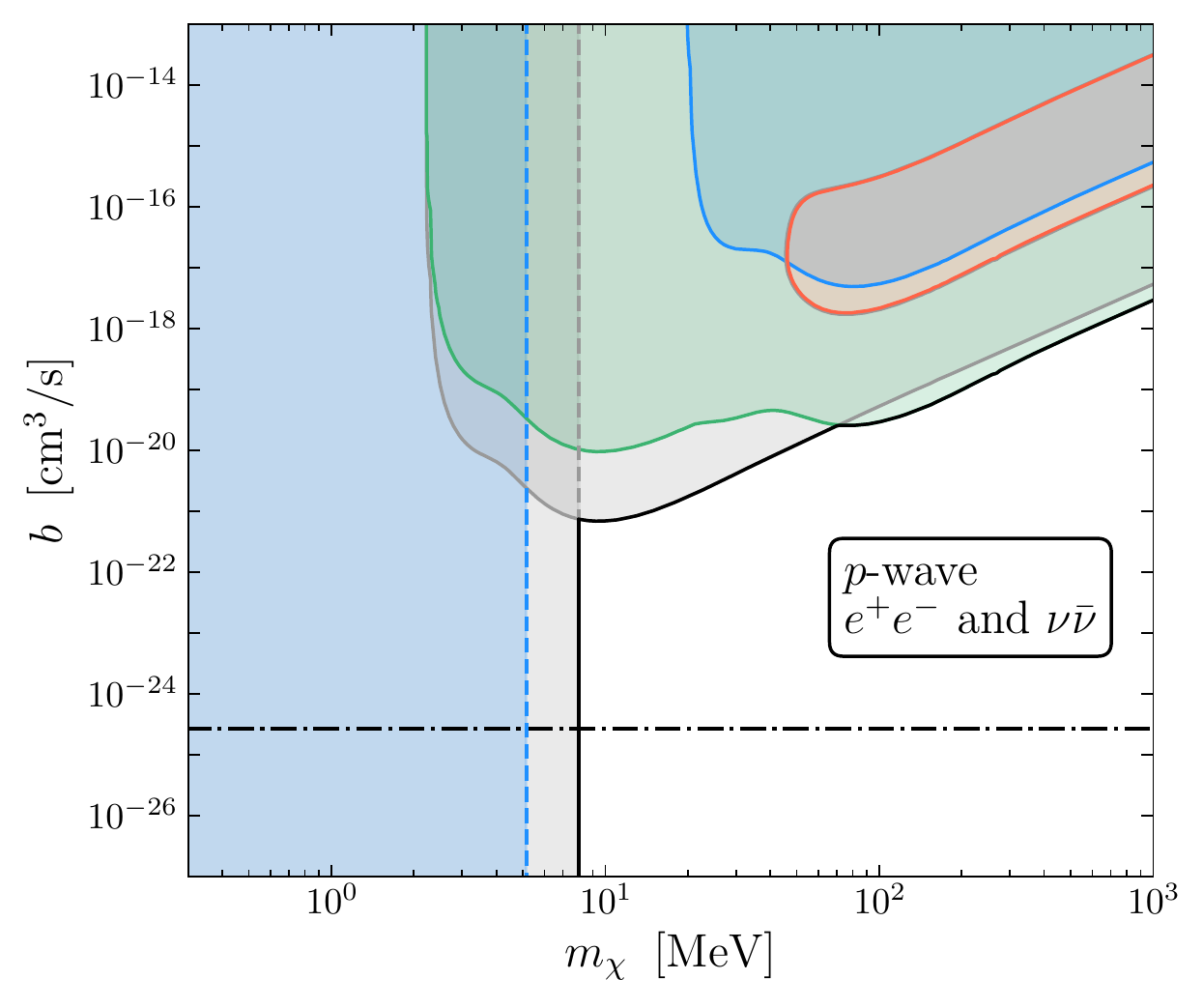}
	\caption{Constraints from BBN (vertical dashed lines, cf.~Fig~\ref{fig:bounds_thermalDMMmass}) and photodisintegration (full lines) for $s$-wave (top panels) and $p$-wave (bottom panels) annihilation of Majorana DM. The left panels correspond to annihilations exclusively into $e^+ e^-$, while the right panels show the results for annihilations into $e^+ e^-$ and $\nu_e \bar{\nu}_e$ with identical cross section. For the $p$-wave scenarios, we show the most stringent limits corresponding to $T^\mathrm{kd} = 100~\mathrm{eV}$. In addition to the combined limit (solid black line), we also separately list the regions of parameter space which are excluded due to deuterium underproduction (gray), deuterium overproduction (orange), $^4\text{He}$ underproduction (blue), $^4\text{He}$ overproduction (pink) and/or $^3\text{He}$ overproduction relative to deuterium (green). For reference, the thermal cross-section is indicated as a black dash-dotted curve.} 
	\label{fig:bounds_photodis_with_mass}
\end{figure}

For a DM particle with a mass satisfying the bounds shown in Fig.~\ref{fig:bounds_thermalDMMmass}, one can still set an upper limit on its annihilation cross section via the constraints on photodisintegration of light nuclei (cf.~section~\ref{sec:pdi}). We show the resulting bounds on the $s$- and $p$-wave coefficient of $\langle \sigma v \rangle \simeq a + b \langle v_\text{rel}^2 \rangle$  in the upper and lower panels of Fig.~\ref{fig:bounds_photodis_with_mass}, respectively. The left panels correspond to the case of DM annihilating exclusively into\footnote{Unlike the bounds from BBN, the bounds from photodisintegration depend on the exact electromagnetic branching ratios of the DM particle. In appendix~\ref{appendix} we provide the relevant bounds for a DM particle that annihilates exclusively into $\gamma\gamma$.} $e^+ e^-$, while the right panels assume annihilations into $e^+ e^-$ and $\nu_e \bar{\nu}_e$ with equal branching ratios, leading to constraints less stringent by a factor of two. As discussed in section~\ref{sec:pdi}, for the scenario of $p$-wave annihilating DM the bounds explicitly depend on the kinetic decoupling temperature, which we have fixed to $T^\text{kd} =100\,$eV in this figure for concreteness. This is at the lower end of values consistent with Lyman-$\alpha$ measurements~\cite{Bringmann:2016ilk}; results for other choices of $T^\text{kd}$ will be discussed below. Furthermore, let us note that in Fig.~\ref{fig:bounds_photodis_with_mass} as well as in all following figures we assume the DM particle to be self-conjugate; the photodisintegration bounds as well as the thermal cross-section for a complex scalar or a Dirac fermion are shifted by a factor of two. 

In the various panels of Fig.~\ref{fig:bounds_photodis_with_mass}, the colored regions enclosed by the solid curves show the range of DM masses and annihilation cross sections which are excluded on the basis of photodisintegration. The vertical dashed curves correspond to the limits on the mass of a thermal Majorana DM particle already shown in Fig.~\ref{fig:bounds_thermalDMMmass}.\footnote{While these bounds are indeed virtually independent of the annihilation cross section for $\langle \sigma v \rangle \gtrsim 10^{-30}\,\text{cm}^3/\text{s}$, they are strictly speaking not valid once the annihilation process of DM is not efficient enough to keep it in equilibrium with the SM heat bath at least down to $T \simeq m_\chi$. Such a scenario would in any case imply a drastically different cosmological history of DM compared to the standard thermal freeze-out scenario, and is not further discussed here.} In the gray (orange) shaded regions the D/$^1$H abundance is too small (too large) compared to the observationally inferred value given in Eq.~(\ref{eq:D_abundance}). The constraints from underproduction of ${}^4$He and overproduction of ${}^3$He relative to deuterium are shown in blue and green, respectively. The impact of the energy thresholds for photodisintegration of D and ${}^4$He are clearly reflected in the upturn of the corresponding bounds at $m_\chi \simeq 2.2\,$MeV and $\simeq 21\,$MeV; for smaller masses, all photons produced in the electromagnetic cascade following the initial injection of an $e^+ e^-$ pair are below the threshold energies of $E_{d(\gamma,p)n}^\text{(th)}$ or $E_{{}^4\text{He}(\gamma,n){}^3\text{He}}^\text{(th)}$, respectively.  For $m_\chi \gtrsim 25\,$MeV, we find that depending on the annihilation cross section of DM, photodisintegration processes can either give rise to an over- or underproduction of deuterium implying a very narrow region between the solid red and gray curves in Fig.~\ref{fig:bounds_photodis_with_mass} consistent with the observed value of D/$^1$H. However, these parts of the parameter space are robustly excluded by the constraints on ${}^3$He/D and $\mathcal{Y}_\text{p}$.

\begin{figure}[h]
	\centering
	\includegraphics[width=0.99\textwidth]{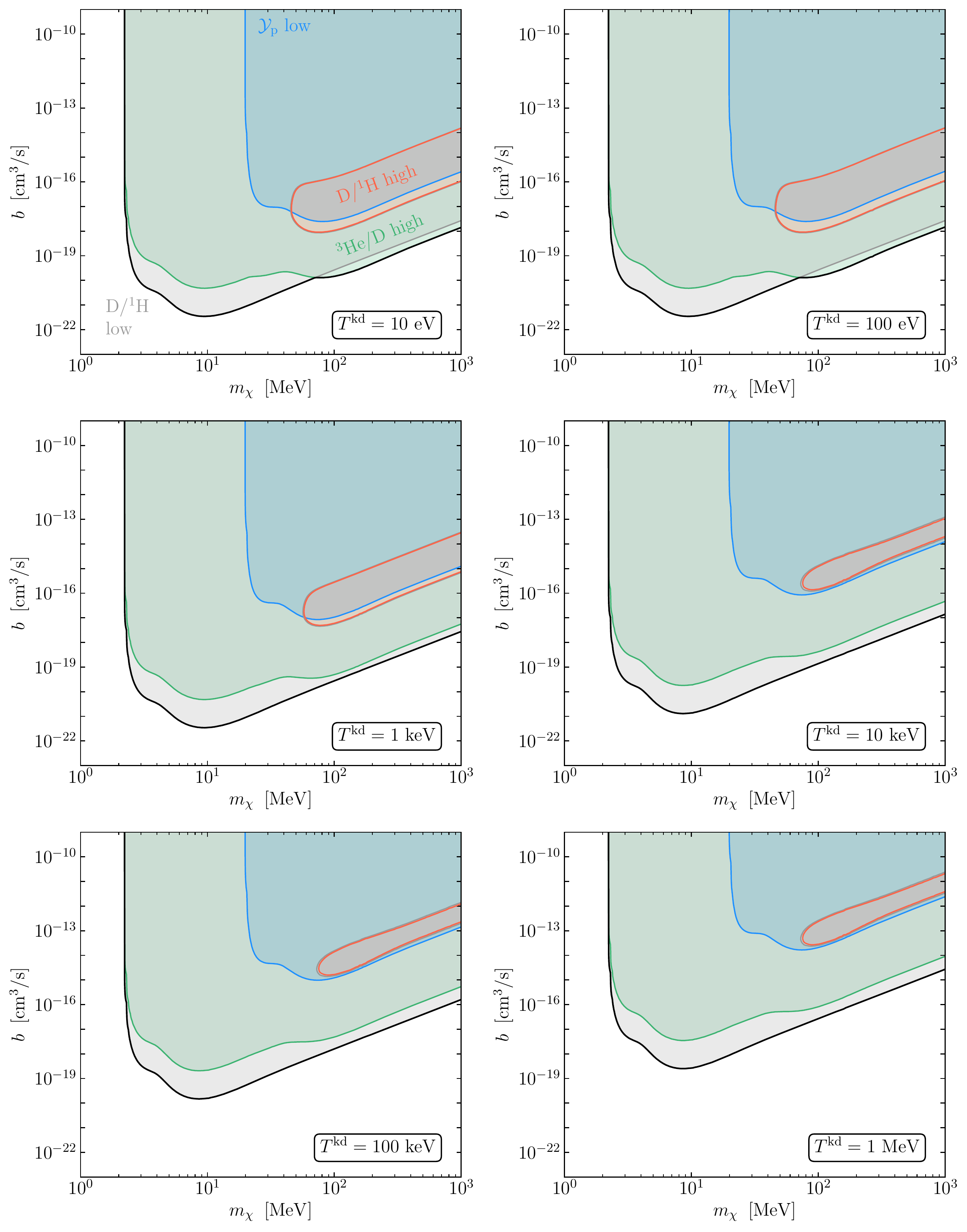}
	\caption{Photodisintegration constraints for the $p$-wave annihilation of self-conjugate DM into electron-positron pairs for six different choices of the kinetic decoupling temperature $T^\text{kd}$. The meaning of the differently colored regions is identical to the one used in Fig.~\ref{fig:bounds_photodis_with_mass}.}
	\vspace{0.3cm}
	\label{fig:bounds_pwave_diff_Tkd}
\end{figure}

For DM masses sufficiently larger than the relevant photodisintegration thresholds, $m_\chi \gtrsim 100\,$MeV, the upper limits on the $s$-wave annihilation cross section $\langle \sigma v \rangle \simeq a$ shown in the top panels of Fig.~\ref{fig:bounds_photodis_with_mass} depend linearly on $m_\chi$. This can be understood from noting that the energy injected into the plasma scales as $E_\text{inj} \propto m_\chi n_\chi^2 a \propto a/m_\chi$.
Conversely, for $p$-wave dominated scenarios the thermally averaged annihilation cross section itself depends on $m_\chi$ via $\langle \sigma v \rangle \simeq b \langle v_\text{rel}^2 \rangle \propto b/m_\chi$ (cf.~Eq.~(\ref{eq:vrel2})), and correspondingly the injected energy depends on the DM mass via $E_\text{inj} \propto 1/m_\chi^2$. This explains the stronger mass dependence of the upper limits on $b$ shown in the lower panels. The strongest bounds from photodisintegration are obtained for $m_\chi \simeq 150\,$MeV ($m_\chi \simeq 10\,$MeV) for the case of $s$-wave ($p$-wave) dominated DM annihilation. In the former scenario, it is almost possible to probe the value of $\langle \sigma v \rangle$ expected for a thermal relic, as indicated by a black dash-dotted line in the various panels of Fig.~\ref{fig:bounds_photodis_with_mass}. 

In the lower panels of Fig.~\ref{fig:bounds_photodis_with_mass} we have assumed that DM kinetically decouples at $T^\text{kd} = 100\,$eV. In order to investigate the dependence of our results on this choice, we show in Fig.~\ref{fig:bounds_pwave_diff_Tkd} the upper limits on $b$ for six different values $T^\text{kd} = 10$ $\text{eV}$, $100\,\text{eV}$, $1\,\text{keV}$, $10\,\text{keV}$, $100\,\text{keV}$, and $1\,$MeV. We observe that for $T^\text{kd} \lesssim 100\,$eV the bound becomes independent of the kinetic decoupling temperature: in this case photodisintegration entirely takes place during the epoch where DM is still in kinetic equilibrium. On the other hand, for $T^\text{kd} \gtrsim 100\,$eV the upper limits on $b$ become less stringent with larger decoupling temperatures, corresponding to smaller values of the thermally averaged velocity square of DM during the times relevant to photodisintegration (cf.~Fig.~\ref{fig:v2DM}). For $T^\mathrm{kd} \gtrsim 10~\mathrm{keV}$ decoupling occurs prior to the onset of photodisintegration, in which case it follows from Eq.~(\ref{eq:TDM_T_R}) that the constraints simply scale as $g_s^{\text{(vis)}} (T^\mathrm{kd})^{2/3} T^\mathrm{kd}$ due to the red-shifting of the DM temperature. A further discussion of this point can be found in appendix~\ref{sec:additional_results}.

\subsection{Comparison with other constraints}
\label{sec:results_comparison}

\begin{figure}[t]
	\centering
	\includegraphics[width=0.495\textwidth]{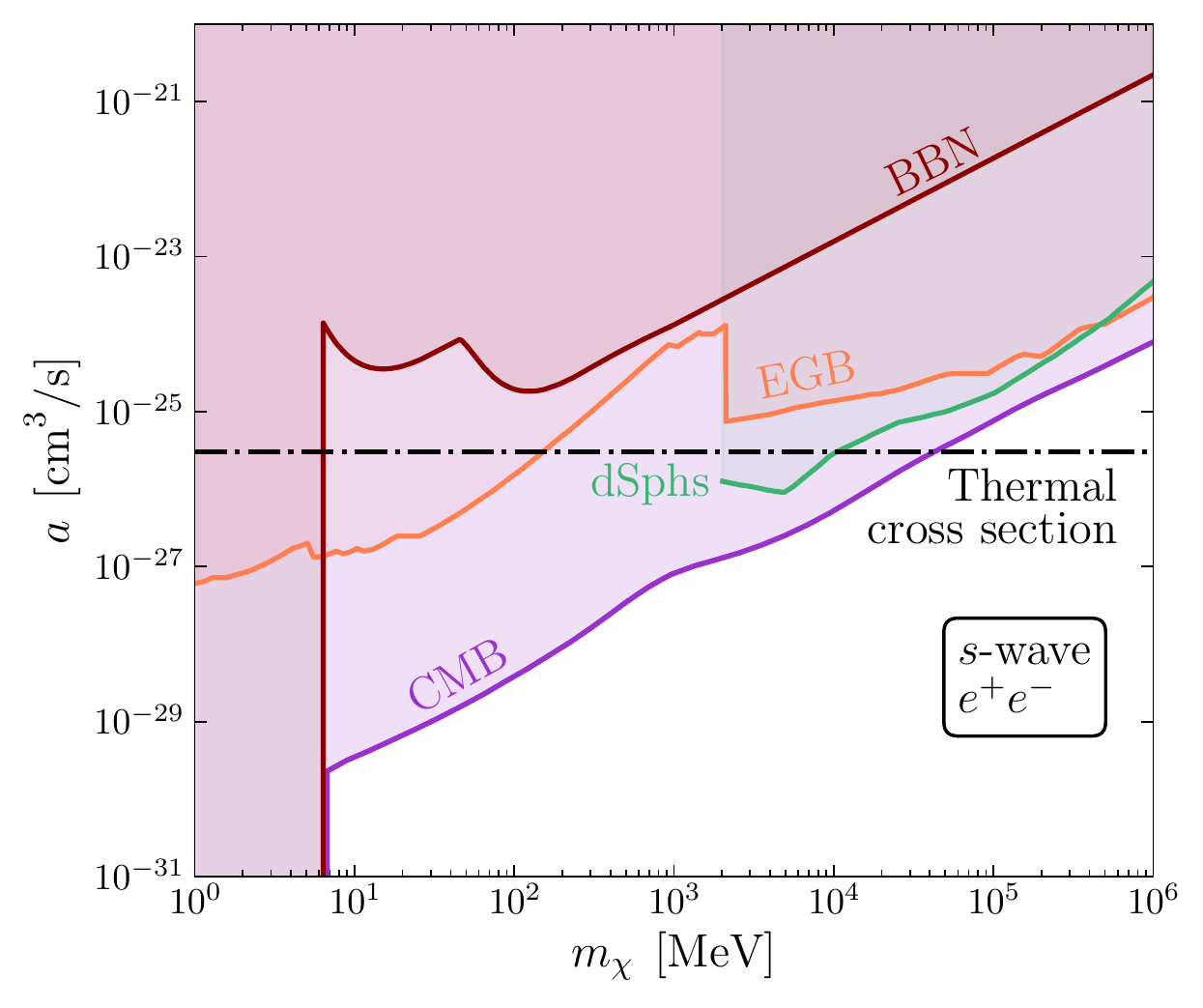}
	\includegraphics[width=0.495\textwidth]{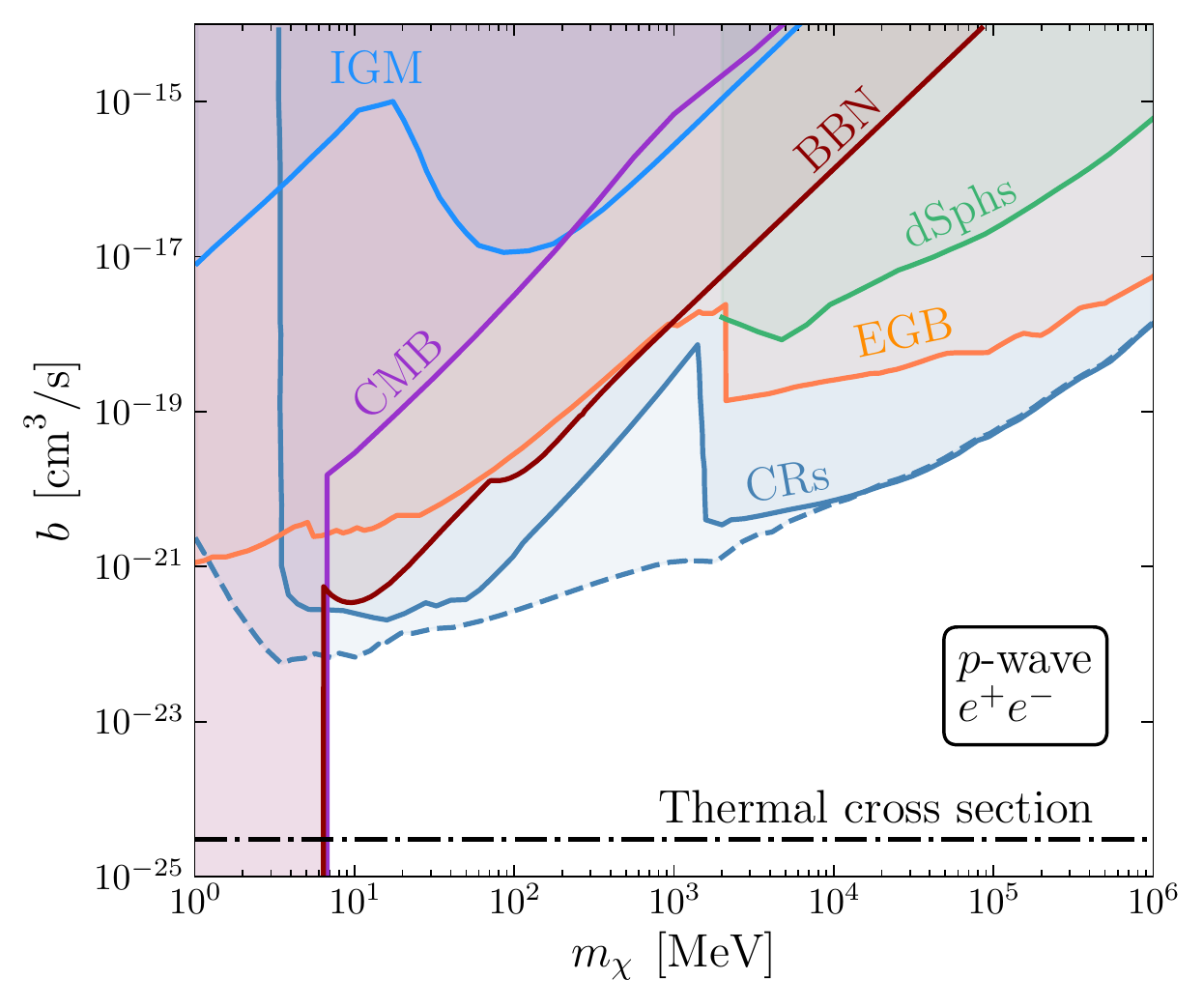}
	\caption{Comparison of the BBN constraints derived in this work (dark red curves) with complementary bounds on the annihilation cross section for $s$- and $p$-wave  annihilations of DM, shown in the left and right panel, respectively. Bounds from the CMB~\cite{Slatyer:2015jla,Liu:2016cnk,Akrami:2018vks} (updated with the results from Fig.~\ref{fig:bounds_thermalDMMmass} for small masses) are shown in purple, from the high-redshift intergalactic medium temperature~\cite{Liu:2016cnk} in orange, from the diffuse extragalactic background~\cite{Essig:2013goa,Massari:2015xea} in red, from gamma-ray observations of MW-satellite dwarf galaxies~\cite{Zhao:2016xie} in green, and from cosmic rays observations~\cite{Boudaud:2018oya} in blue. See text for details.}
	\label{fig:results_comparison}
\end{figure}

Finally, in Fig.~\ref{fig:results_comparison} we compare our constraints on the annihilation cross section of DM into $e^+ e^-$ with bounds from complementary cosmological and astrophysical observations, choosing for definiteness $T^\text{kd} = 100\,$eV. Upper limits on the $s$- and $p$-wave annihilation cross section are shown in the left and right panel, respectively.

In the $s$-wave case, constraints from the CMB on exotic energy injection~\cite{Slatyer:2015jla,Liu:2016cnk,Akrami:2018vks} provide the most stringent results (for $m_\chi \gtrsim 10\,$MeV), probing values of $\langle \sigma v \rangle \simeq a$ at least two orders of magnitude smaller than the bounds on primordial abundances of light nuclei derived in this work. Depending on the DM mass, constraints from the diffuse extragalactic background~\cite{Essig:2013goa,Massari:2015xea} or from gamma-ray observations of Milky Way dwarf satellite galaxies~\cite{Zhao:2016xie} are competitive or more stringent than the bounds from BBN. 

On the other hand, for $p$-wave dominated DM annihilations the bounds from photodisintegration derived in this study are highly competitive with all other searches. For all ranges of DM masses, the corresponding upper limit on $b$ is more stringent than the one derived from CMB observations by at least one order of magnitude.\footnote{Details on how we extract complementary bounds on the $p$-wave annihilation cross section from results in the literature can be found in appendix~\ref{app:pwavebounds}.} As explained above, this is mainly due to the significantly larger velocities of DM during the times relevant to photodisintegration compared to recombination, cf.~Fig.~\ref{fig:v2DM}. Recently, it has been pointed out that the observations of charged cosmic rays can also be used to constrain $p$-wave annihilating DM at the MeV scale~\cite{Boudaud:2018oya}; taken at face value, those bounds are a factor of a few more stringent than the bounds from BBN. However, considering the significant systematic uncertainties from cosmic ray propagation inherent to those constraints (with an example of two different propagation models shown via the solid and dashed blue curves in the right panel of Fig.~\ref{fig:results_comparison}), the results from BBN derived in this work clearly serve as an important complementary constraint.

\section{Conclusions}
\label{sec:conclusions}

Stable particles at the MeV scale constitute a theoretically interesting and phenomenologically rich class of candidates for dark matter. If they interact sufficiently strong with SM particles, they will be in thermal equilibrium at early times, and eventually obtain their relic abundance via the well-known freeze-out mechanism. Besides being testable by direct and astrophysical searches, such particles can also leave their imprint on cosmological observables.

In the first part of this study, we have revisited the lower bound on the mass of thermal dark matter from BBN and CMB observations, employing the most recent set of available data. In addition to increasing the expansion rate of the Universe, the annihilation of MeV-scale dark matter generically leads to a modification of the photon and/or neutrino temperature via the injection of such particles into the thermal bath. We also show that dark matter can serve as a mechanism to keep electrons, positrons, and neutrinos in equilibrium after the relevant SM processes stop being efficient, leading to a modification of the effective number of neutrinos $N_\text{eff}$. We then derive constraints for the two scenarios which are most natural from the particle physics point of view, namely annihilations exclusively into $e^+ e^-$, or into $e^+ e^-$ and $\nu \bar \nu$ with similar branching ratios. BBN bounds on the latter scenario are discussed for the first time in this work. 

Our results imply that BBN and CMB independently exclude a thermal dark matter particle with a mass below $\simeq (7-10)\,$MeV, with the detailed value depending on the dark matter particle type and annihilation channel. This lower bound serves as an important target value for future efforts in the search for low-mass dark matter, conducted e.g.~by direct detection or beam dump experiments. Importantly, the constraints from BBN derived in this work are quite robust with respect to modifications of the particle content in the dark sector, while the bound from the CMB at least in some scenarios can be circumvented by postulating the existence of additional species in the dark sector.

Even for masses above $\simeq 10\,$MeV, BBN constitutes an important probe for the annihilation cross section of dark matter. Residual annihilations into SM particles induce an electromagnetic cascade on the background photons, electrons and nuclei, which in turn can lead to photodisintegration of deuterium and helium produced earlier during nucleosynthesis. We have extended previous works in this direction by considering for the first time dark matter masses below $\lesssim 100\,$MeV, where the breakdown of the universal spectrum of cascade photons as well as the onset of photodisintegration thresholds requires a dedicated study. For the case of $s$-wave annihilations, we find that the constraints from the CMB on exotic energy injection provides stronger limits than the one derived from primordial nuclear abundances. In contrast, $p$-wave annihilating dark matter is only mildly constrained by CMB data due to the very small velocities during recombination. On the other hand, the suppression of the annihilation cross section by $\langle v_\mathrm{rel}^2 \rangle$ is much less severe during the times relevant to photodisintegration, corresponding to temperatures $T \simeq 1\,\text{keV}\gg T_\text{CMB}$. Consequently, we find that BBN imposes significantly more stringent bounds on the annihilation cross section for $p$-wave suppressed scenarios, which are furthermore competitive with complementary bounds derived from observations of cosmic rays. 

\acknowledgments

We thank Camilo Garcia-Cely and Felix Kahlhoefer for useful discussions. This work is supported by the ERC Starting Grant `NewAve' (638528) as well as by the Deutsche Forschungsgemeinschaft under Germany‘s Excellence Strategy -- EXC 2121 `Quantum Universe' -- 390833306.

\pagebreak
\appendix
\section{Compilation of additional results}
\label{appendix}
\label{sec:additional_results}

In this appendix, we provide additional results for the upper limits on the annihilation cross section of DM originating from photodisintegration (see also section~\ref{sec:results_pdi}, in particular Figs.~\ref{fig:bounds_photodis_with_mass} and~\ref{fig:bounds_pwave_diff_Tkd}).

\begin{figure}[h]
	\centering
	\includegraphics[width=0.99\textwidth]{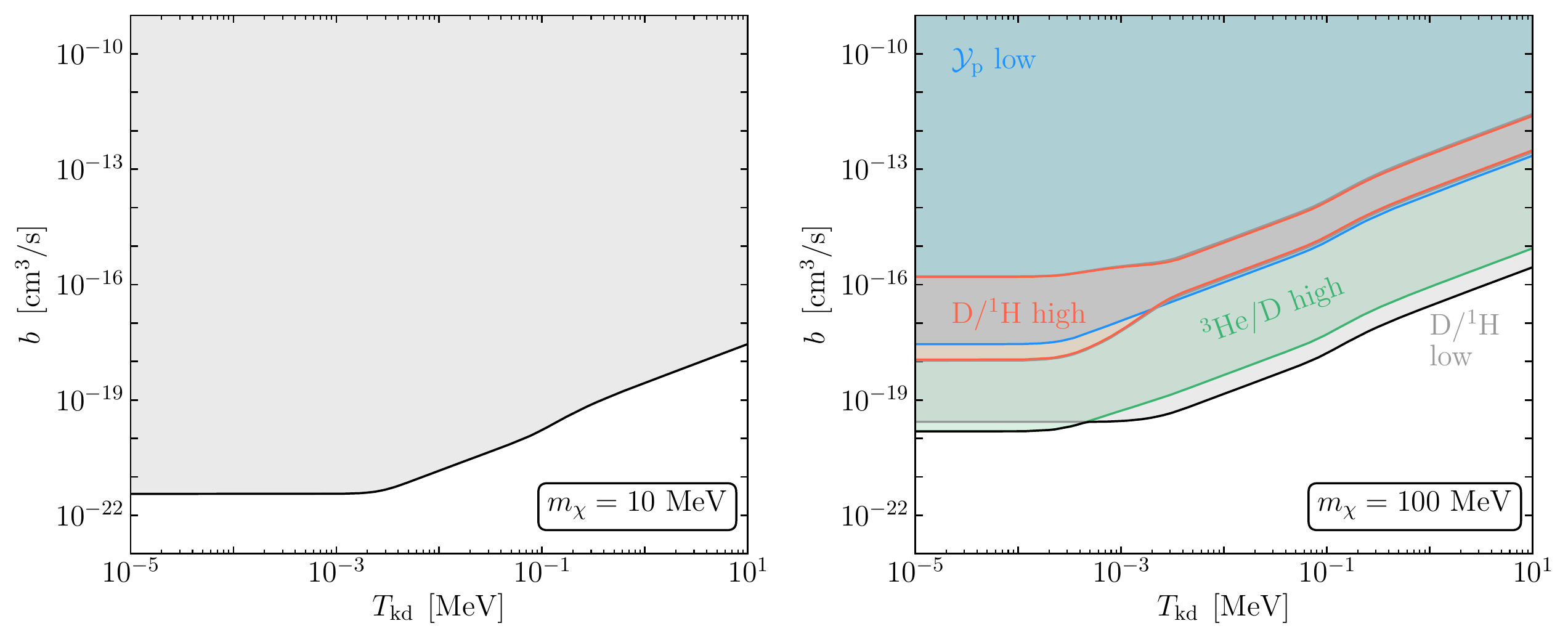}
	\caption{Photodisintegration constraints for the $p$-wave annihilation of a self-conjugate DM particle into electron-positron pairs as a function of the kinetic decoupling temperature $T^\text{kd}$ for two different DM masses $m_\DM$. The combined limit is shown as a solid black line, while the individual contributions from under- or overproduction of deuterium are shown in gray and orange, from underproduction of $^4\text{He}$ in blue, and from overproduction of $^3\text{He}$ relative to deuterium in green.}
	\label{fig:dep_Tkd}
\end{figure}

In Fig.~\ref{fig:dep_Tkd}, we investigate the dependence of our bounds on the kinetic decoupling temperature $T^\text{kd}$ of DM by showing the upper limit on the coefficient $b$ of the $p$-wave annihilation cross section as a function of $T^\text{kd}$. In the left (right) panel, the mass of the DM particle is fixed to $m_\DM = 10\,\mathrm{MeV}$ ($m_\DM = 100\,\mathrm{MeV}$), which roughly corresponds to the global (local) minimum of the exclusion limits shown in Fig.~\ref{fig:bounds_pwave_diff_Tkd}. As discussed in section~\ref{sec:results}, the bounds become independent of $T^\text{kd}$ for small values of the kinetic decoupling temperature, as at some point the DM particles decouple only after the time relevant for photodisintegration. The value of $T^\text{kd}$ at which this happens scales approximately with the inverse of the DM mass and is given by $T^\text{kd, min} \sim \mathcal{O}(1\,\mathrm{keV})$ and $T^\text{kd, min} \sim \mathcal{O}(100\,\mathrm{eV})$ for $m_\DM = 10\,\mathrm{MeV}$ and $m_\DM = 100\,\mathrm{MeV}$, respectively. On the other hand, it follows from Fig.~\ref{fig:dep_Tkd} (and also from Fig.~\ref{fig:bounds_pwave_diff_Tkd}) that for large values of $T^\text{kd}$, the decoupling of the DM particles happens before photodisintegration, implying smaller velocities and therefore weaker constraints on the annihilation cross section. As mentioned previously, the corresponding bounds then scale as $g_s^{\text{(vis)}} (T^\mathrm{kd})^{2/3} T^\mathrm{kd}$. For $T^\mathrm{kd} \lesssim m_e$ this can be noticed in a slight bump in the constraint due to the onset of electron-positron annihilation.

\begin{figure}[h]
	\centering
	\includegraphics[width=0.495\textwidth]{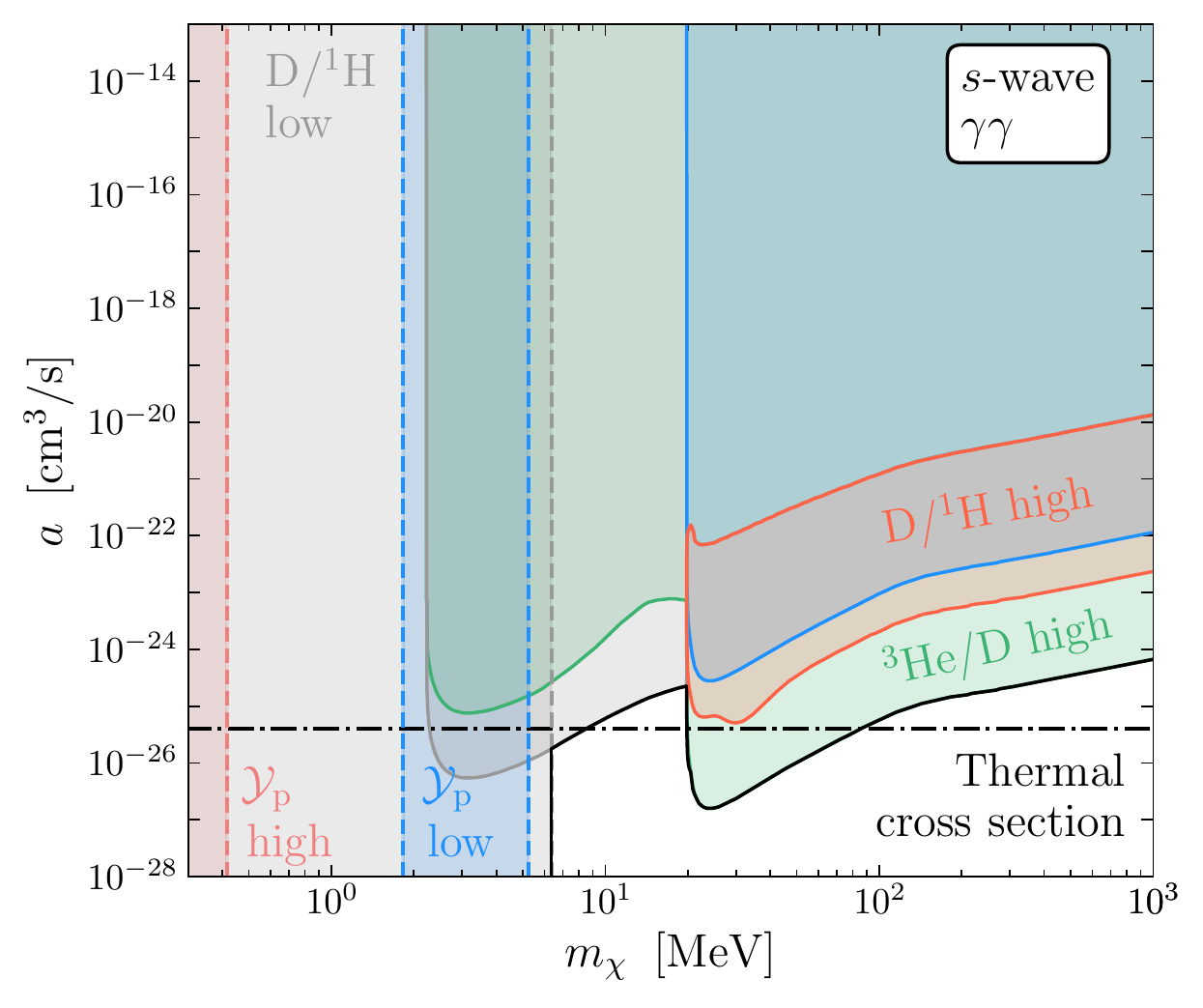}
	\includegraphics[width=0.495\textwidth]{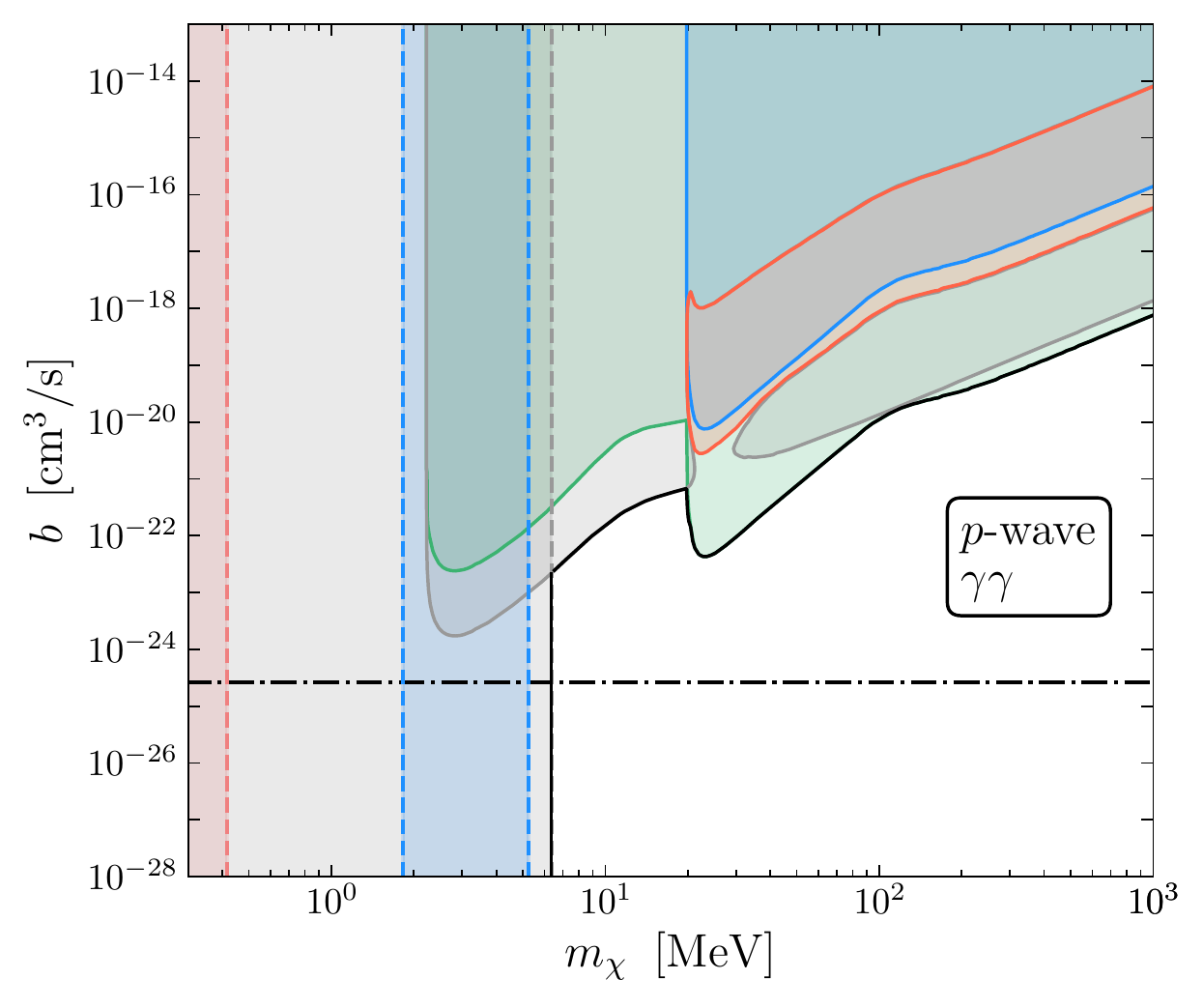}
	\caption{Constraints from BBN (vertical dashed lines, cf. fig.~\ref{fig:bounds_thermalDMMmass}) and photodisintegration (full lines) for $s$-wave (left panel) and $p$-wave (right panel) annihilation of Majorana DM into a pair of photons. For the $p$-wave scenario, we show the most stringent limit corresponding to a kinetic decoupling temperature $T^\mathrm{kd} = 100~\mathrm{eV}$. In addition to the combined limit (solid black line), we also separately list the regions of parameter space which are excluded due to deuterium underproduction (gray), deuterium overproduction (orange), $^4\text{He}$ underproduction (blue), $^4\text{He}$ overproduction (pink) and/or $^3\text{He}$ overproduction relative to deuterium (green). The thermal cross-section is shown for reference as a black dash-dotted curve.}
	\label{fig:bounds_photon}
\end{figure}

\begin{figure}[t]
	\centering
	\includegraphics[width=0.99\textwidth]{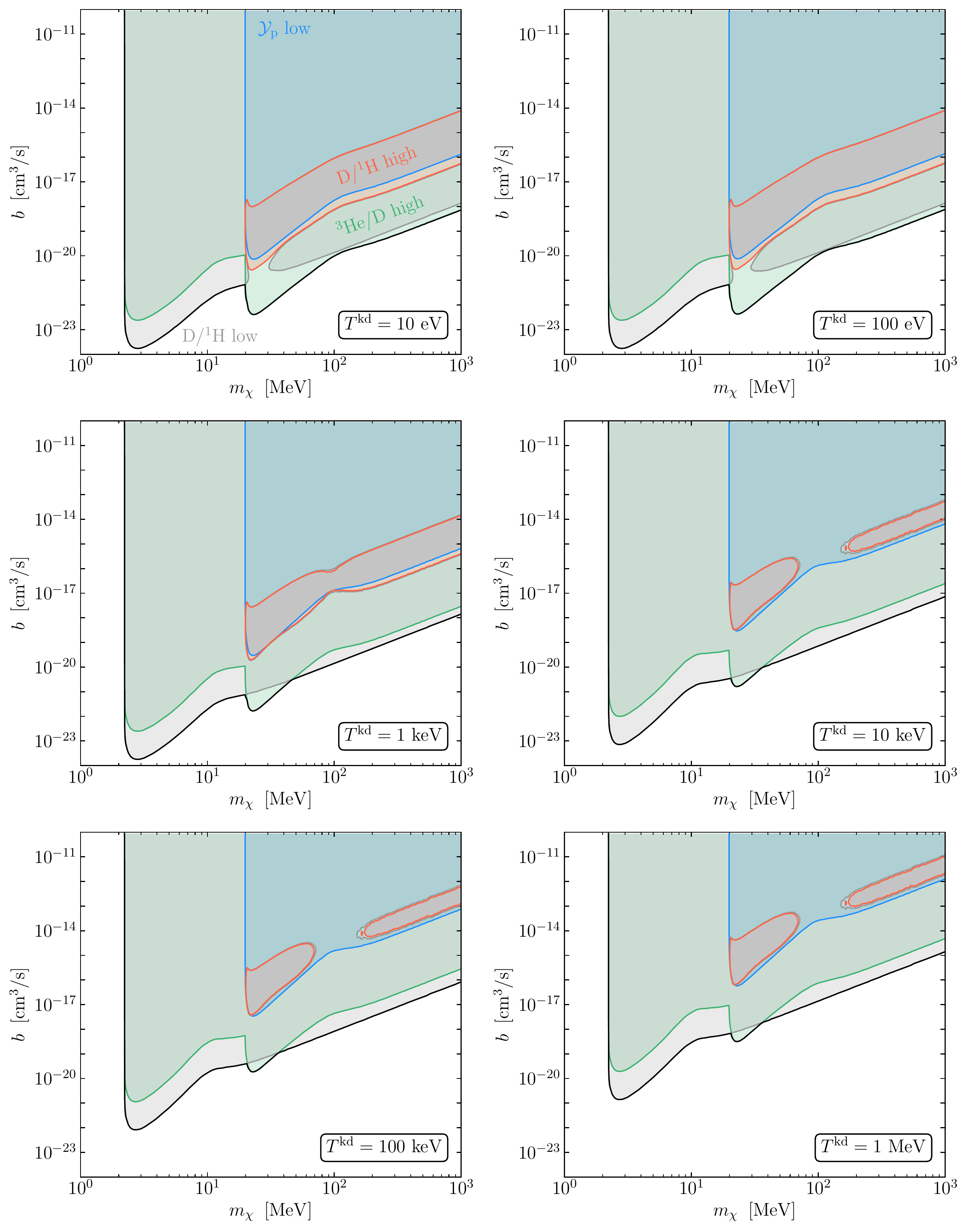}
	\caption{Photodisintegration constraints for the $p$-wave annihilation of a Majorana DM particle into a pair of photons for different choices of the kinetic decoupling temperature $T^\text{kd}$. The color coding is identical to the one used in Fig.~\ref{fig:bounds_photon}.}
	\label{fig:bounds_photons_Tkd}
	\hspace{2cm}
\end{figure}

When presenting our main results in Figs.~\ref{fig:bounds_photodis_with_mass} and~\ref{fig:bounds_pwave_diff_Tkd}, we have assumed that DM either annihilates exclusively into $e^+ e^-$ or into $e^+ e^-$ and $\nu_e \bar{\nu}_e$ with identical cross section. While this is indeed well-motivated from the point of view of particle physics, in principle DM could also dominantly annihilate into a pair of photons. In light of this, we show in Fig.~\ref{fig:bounds_photon} the limits from photodisintegration as well as the lower bound on the mass of a thermal DM particle from BBN for a scenario with annihilations exclusively into two photons (in analogy to Fig.~\ref{fig:bounds_photodis_with_mass}). For the case of $p$-wave annihilation, we again set $T^\text{kd} = 100\,\mathrm{eV}$ for definiteness. As expected, in this case one obtains substantially stronger bounds on the annihilation cross section compared to the case of annihilations into electron-positron pairs: in the latter scenario, the spectrum of photons originating from the electromagnetic cascade is softer, as only electrons and positrons are produced as primary particles in the DM annihilation, while in the former case the photons also originate directly from DM.
Finally, for completeness we present in Fig.~\ref{fig:bounds_photons_Tkd} the photodisintegration bounds for annihilations into $\gamma \gamma$ for different values of the kinetic decoupling temperature, in analogy to Fig.~\ref{fig:bounds_pwave_diff_Tkd}.


\pagebreak
\section{Complementary bounds on the $p$-wave annihilation cross section}
\label{app:pwavebounds}

We estimate CMB constraints on the $p$-wave annihilation cross section of DM based on the $s$-wave bounds derived in~\cite{Liu:2016cnk} as follows: first, the constraints are rescaled with a factor of $3.2/4.1$ to account for the updated calculation of $p_\mathrm{ann}$~\cite{Aghanim:2018eyx}. We then obtain a bound on the $p$-wave coefficient of the thermally averaged annihilation cross section via
\begin{align}
b = \frac{a(m_\DM)}{\left< v_\mathrm{rel}^2 \right> (z = 600)} = a(m_\DM) \frac{m_\DM T^\mathrm{kd}}{6 T (z = 600)}\,,
\end{align}
with $T (z = 600) \simeq 0.14\,\mathrm{eV}$, the $s$-wave bound $a(m_\DM)$ from~\cite{Liu:2016cnk}, and $T^\mathrm{kd} = 100~\mathrm{eV}$ in accordance with the value chosen for our BBN bound. Here we used the fact that the CMB constraints are most sensitive around $z\sim 600$. Finally, we combine this constraint with the lower bound on the mass from Fig.~\ref{fig:bounds_thermalDMMmass}.

The remaining $p$-wave constraints from the high-redshift intergalactic medium temperature (IGM)~\cite{Liu:2016cnk}, the diffuse extragalactic background (EGB)~\cite{Essig:2013goa,Massari:2015xea}, and gamma-ray observations of MW-satellite dwarf galaxies (dSphs)~\cite{Zhao:2016xie} are all given in terms of $b v^2$ with some reference velocity $v$. In order to obtain a bound on $b$, we therefore have to rescale the results with the corresponding value of $v$, which are given by $v = 100\,\text{km}/\text{s}$ (IGM), $v = 220\,\text{km}/\text{s}$ (EGB), $v = 270\,\text{km}/\text{s}$ (dSphs), respectively. In addition to this, we divide the dSphs bounds by a factor of six due to a different normalization convention chosen in~\cite{Zhao:2016xie}.

We furthermore note the possibility to constrain $p$-wave annihilations using the steep density profiles and correspondingly enhanced DM velocities in the vicinity of supermassive black holes~\cite{Shelton:2015aqa}, which however is subject to significant astrophysical uncertainties.
\bibliography{refs}
\bibliographystyle{ArXiv}

\end{document}